# Error Detection and Correction for Distributed Group Key Agreement Protocol

P.Vijayakumar[1], S.Bose[1], A.Kannan[2]

[1] Department of Computer Science & Engineering, Anna University, Chennai, India -600025
vijibond2000@gmail.com, sbs@cs.annauniv.edu
[2] Department of Information Science & Technology, Anna University, Chennai, India-600025
kannan@annauniv.edu

## Abstract

*Integrating an efficient Error detection and correction scheme with less encoding and decoding complexity to support the distribution of keying material in a secure group communication is an important issue, since the amount of information carried out in the wireless channel is high which produces more errors due to noise available in the communication channel. Moreover, the key must be sent securely to the group members. In this paper, we propose a new efficient group key computation protocol that provides more security and also integrates an encoding method in sender side and decoding method in the receiver side. To achieve security in key computation process, we propose Euler's totient function based Diffie-hellman key distribution protocol. To provide efficient error detection and correction method while distributing the Keying and re-keying information, we introduce tanner graph based encoding stopping set construction algorithm in sender and receiver side of the group communication. Two major operations in this scheme are joining and leaving operations for managing group memberships. The encoding and decoding complexity of this approach is computed in this paper and it is proved that this proposed approach takes less decoding time complexity.*

## 1 INTRODUCTION

Wireless multimedia services such as videoconferences, sporting events, audio and video broadcasting are based upon Group communication where multimedia messages are shared to a group of members with less computation, communication cost due to the limitation of battery power. In such a scenario only registered members of a group can receive multimedia data. Group can be classified into static and dynamic groups. In static groups, membership of the group is predetermined and does not change during the communication. In dynamic groups, membership can change during multicast group communication. When a new member joins into the service, it is the responsibility of the Group Centre (GC) to disallow new members from having access to previous data. This provides backward secrecy in a secure multimedia communication. Similarly, when an existing group member leaves from any group, he/she do not have access to future data. This achieves forward secrecy. GC also takes care of the job of distributing the Secret key and Group key to group members. Therefore, in dynamic group communication, members may join or depart from the service at any time. The number of keys to be updated is high when there is a change in group membership. Moreover all those keys are needed to be communicated to the group members with minimal computation and communication time. In this paper, we propose a new group key Distribution scheme based on Eulers totient function that reduces the computation time compared with other existing approaches.

Most basic key distribution schemes mainly focuses on the domain of key computation which aims at reducing the storage and computation complexity. However, some of the literatures focus on packet loss and packet recovery in turn. In this paper we propose a new key distribution protocol along with error detection and correction techniques. Hence, this approach provides a good way for the group members to construct the original key even if the keying/Rekeying information's that are sent through the wireless channel are lost. The remainder of this paper is organized as follows: Section 2 provides the features of some of the related works. Section 3 discuses the overview of key distribution protocol. Section 4 provides the detailed explanation of the proposed Eulers Totient Function (ETF) based key computation work. Section 5 shows the performance results. Section 6 discuses the integration of Error detection and correction to our proposed key computation method. Section 7 gives the concluding remarks and suggests a few possible future enhancements.

## 2    LITERATURE SURVEY

There are many related works on key management and key distribution that are present in the literature [1-3], [10, 14-16] that can be divided into two types of group key management namely centralized and distributed key management scheme. In Centralized key management scheme, entire key generation and computation is performed by a single entity known as Group Centre (GC). A Special case is the scenario where the key is generated by some Trusted Third Party (TTP) which is not a group member [4, 8]. This type of key management is called as distributed key management scheme. In distributed key management scheme, each group member makes an independent contribution to the group key. We make a further distinction among two slightly different flavors of contributory key agreement namely Partially Contributory and Fully Contributory. In Partially Contributory key management, some operations result in contributory and others in centralized key agreement. In Fully Contributory key management, all key management operations are contributed by each group member. Centralized key agreement is the most intuitive and the most natural. It has been used in a number of past and current mechanisms and its use is commensurate with important advantages as well as certain drawbacks. One such drawback is the overall reliance on a single party. Among the various works on Centralized key distribution, Maximum Distance Separable (MDS) [5] method focuses on error control coding techniques for distributing re-keying information. In MDS, the key is obtained based on the use of Erasure decoding functions [6] to compute session keys by the GC/group members. The main limitation of this scheme is that it increases both computation and storage complexity since it uses more parameters. The computational complexity is obtained by formulating $l_r+(n-1)m$ where $l_r$ is the size of r bit random number used in the scheme and m is the number of message symbols to be sent from the group center to group members. If $l_r=m=l$, computation complexity is *nl*. The storage complexity is given by $\lceil log_2 L \rceil + t$ bits for each member. L is number of levels of the Key tree. Hence Group Center has to store $n(\lceil log_2 L \rceil + t)$ bits. Wade Trappe, et al proposed a Parametric One Way Function (POWF) [7] based binary tree Key Management. The storage complexity is given by $(log_\tau n) + 2$ keys for a group centre. The amount of storage needed by the individual user is given as $\frac{(\tau^{L+1}-1)}{\tau-1}$ keys. Computation time is represented in terms of number of multiplication required. The number of multiplication needed to update the KEKs using bottom up approach *is* $\tau log_\tau n - 1$. Multiplication needed to update the KEKs using top down approach is $\frac{(\tau-1)log_\tau n(log_\tau n+1)}{2}$. Here $\tau$ represents the degree of the tree. In the domain of group communication, contributory key agreement has been, for the most part, restricted to the cryptographic literature [8, 13-17]. A new group keying method that uses one-way functions [8] to compute a tree of keys, called the One-way Function Tree (OFT). In this method, the keys are computed up the tree, from the leaves to the root. This approach reduces re-keying broadcasts to only about *log n* keys. The major limitation of this approach is that it consumes more space. However, time complexity is more important than space complexity.

    The scheme MABS-B [11] provides perfect resilience against packet loss by eliminating the correlation among the packets that are sent. For ensuring such a scheme, Merkle tree which is based on hash functions is constructed and found to be very efficient for providing batch signature and verification. Meanwhile, due to the limitations in MABS-B, an efficient method for multicast communication with forward security, ForwardDiffSig was proposed [20]. This scheme was found to be very efficient in terms

of speed, exhibiting low delay even for long keys. The proposed contribution of this work is that a variation of LDPC (Low Density Parity Check) error correction codes [18], [21-24]. LDPC is an error correcting code that constructs a parity check matrix M, which is multiplied with the original data words, d to provide a list of code words, c. If the original data word consists of 8 bits, then LDPC (8, 16) parity check matrix is generated. LDPC codes can also be described by their parity check matrix [25] or tanner graph. So the degree of the bit node in a tanner graph is equivalent to the column weight of the corresponding column of the parity check matrix. Different Column of a parity matrix will have different column weights. Different row of a matrix will have different row weights. Initially, Tanner graphs [26] were developed for the process of decoding using LDPC codes, in fact, they can be used for the encoding of LDPC codes [12] In order to provide efficient error correction, and we make use of the idea of Tanner graphs. The Tanner graph may produce pseudo tree [19], based encoding stopping set [18]. In the proposed algorithm, the time complexity of error correction procedure is significantly minimized and the proof is given in section 6.2. In this paper we propose a new fully contributed binary tree based key management scheme using Euler's Totient Function $\varphi(n)$ [9]. We have also compared the result obtained from this approach with the previously proposed key computation protocols [14], [16-17]. From the results it is clearly evident that our proposed algorithm reduces computation time. We also Integrates error detection and correction algorithm both in sender and receiver side for the proposed group key computation protocol.

# 3 PROPOSED KEY COMPUTATION PROTOCOL

The proposed framework works in three phases. The first phase is the Group Initialization, where the multiplicative group is created. In the second phase of Member Initial Join, the members send the joining request to the existing group members and obtain all the necessary keys for participation. The final phase of Rekeying deals with all the operations to be performed after a member leaves/joins from the group (providing forward/backward secrecy).

## 3.1 Group Initialization

Initially, the group members select a large prime number p. This value, p helps in defining a multiplicative group $z_p^*$ and a secure one-way hash function H(.). The defined function, H(.) is a hash function defined from $X \times Y = Z$ where X and Y are non-identity elements of $z_p^*$. Since the function H(.) is a one way hash function, x is computationally difficult to determine from the given function $Z = y^x$ (mod p) and y.

## 3.2 Member Initial Join

Whenever a new user 'i' is authorized to join in a group for the first time, the user selects a secret key $K_i$ from the group $z_p^*$, which is known only to the user $U_i$ who computes the Euler's Totient Function value of it. The result is represented as $x = \varphi(K_i)$ which is used as a component in secure one way hash function. Next, it computes the Public key by using the parameter p (group Size) and a value y which is selected from the group $z_p^*$ such that y < p. New user 'i' sends join request along with its public key to the entire remaining user's and also gets all users public key for computing the group key.

## 3.3 Rekeying

Whenever some new member join or some old member leave the group, the existing group members need to compute the new Group Key (GK) in such a way that all the existing members should have the same group key. In such computational scenario, the new group key should be computed in minimal computation time. During the key computation process one node will be designated as a support node, where this node will usually be located nearest to the member leave/join node. If the tree is unbalanced the support node will be located in shallowest right most area. If the tree is a balanced one, any node can become a support node.

## 4 KEY COMPUTATION PROTOCOL (ETF)

In distributed key management environment, the GC is not responsible for computing the GK and SGK. Each member is generating GK via each user's and internal nodes public key. Each member $M_i$ holds a pairs of keys called Secret Key (SK) and Public Key (PK). They notations used to represent the secret and public key are $K_{Mi}$ (*the secret key of member $M_i$*) and $PK_{Mi} = Y^{X_{Mi}} \bmod p$ (*the Public key of member $M_i$*), which will remain valid from the time $M_i$ joins until it leaves. With the help of each user's Public keys a group key is computed when a member join or leave from the service. Group key can be used to encrypt and decrypt the data that is shared between the group members. In this key management scheme a binary key tree is formed in which each node v represents a secret (private) key $K_v$ and a Public Key $PK_v$. Public key can be calculated by using the function $PK_v = y^{\varphi(K_v)} \bmod p$ where y and p are public parameters for that group. The function $\varphi(K_v)$ represents euler's totient value of the secret key $K_v$. Every member holds the secret keys along the key path from his leaf node to the root node.

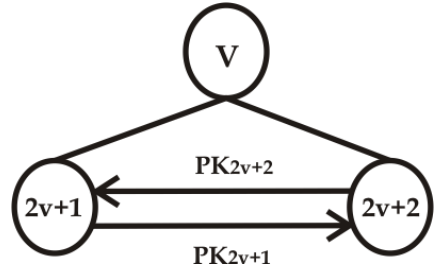

**Fig.1. Calculation of a node value**

For simplicity, we assume that each member knows the public keys of all other group members who are in the key tree. Initially, each member randomly selects the secret key of a leaf node. The secret key of a non-leaf node v can be generated as shown in Fig.1.

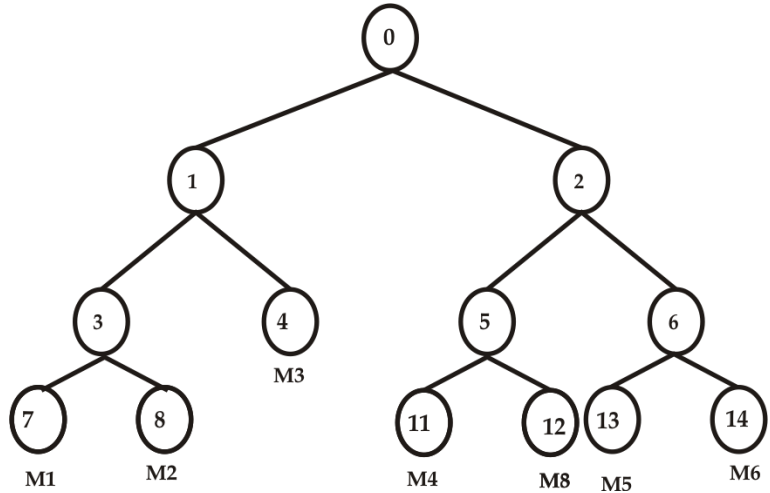

**Fig.2. Binary tree Key management scheme**

Since the member 2v+1 knows the Public Key of member 2v+2 the member 2v+1 can calculate the value of node v by,

$$GK_V = PK_{2v+2}{}^{\varphi(K_{2v+1})} = \left(y^{\varphi(K_{2v+2})}\right)^{\varphi(K_{2v+1})} \bmod p \qquad (1)$$

Similarly member 2v+2 knows the public key of 2v+1, this member can compute the node value by,

$$GK_V = PK_{2v+1}{}^{\varphi(K_{2v+2})} = \left(y^{\varphi(K_{2v+1})}\right)^{\varphi(K_{2v+2})} mod\ p \qquad (2)$$

The computed values shown in equation (1) and equation (2) should be same. Each user can generate GK via all others and intermediate nodes Public key. For example in Fig.2 the member $M_1$ can generate group key via the following steps:
- Using $K_7$ and $PK_8$, the node key $K_3$ is calculated
- After computing $K_3$, $K_3$ and Public key $PK_4$ are used to calculate the node key $K_1$
- Finally, using $K_1$ and $PK_2$ the root key $K_0$ (Group Key) is calculated

The same procedure is used by all other members of the group for computing the GK when there is a change in group membership.

### 4.1    Member joins

Consider a binary tree depicted in Fig.3 that has n members {$M_1$, $M_2$…$M_n$}. The new member $M_{n+1}$ initiate the protocol by broadcasting a join request message that contains its own Public Key $PK_{n+1}$. This message is distinct from any JOIN messages generated by the underlying communication system. Each current member receives this message and first determines the insertion point in the tree. The insertion point is the shallowest rightmost node, where the join does not increase the height of the key tree. The member which is located in that insertion point becomes a support node. Otherwise, if the key tree is fully balanced, any of the leaf nodes can act as *support node* to insert the new member in the key tree structure. The support node has to find the insertion point for the new member. After finding the insertion point, the support node creates a new intermediate node, a new member node, and promotes the new intermediate node to be the parent of both the insertion node and the new member node. The support node is responsible for updating all the internal node keys located in the path from leaf node to the root node. After the updation process, the support node broadcasts the public key of updated key nodes to essential group members. On reception of the public keys, all other members in the key tree update their group key. Only the required public keys for the computation of group key are sent to the group members, since all the other keys are already known to them and it might appear to increase the network traffic. Fig.3 (a) and (b) illustrates the case of member join/member leave. Suppose if member M8 want to join in this group then the keys from the leaf node to the root node must be updated in order to provide backward secrecy. First, the new joining user broadcasts its public key PK12 on joining.

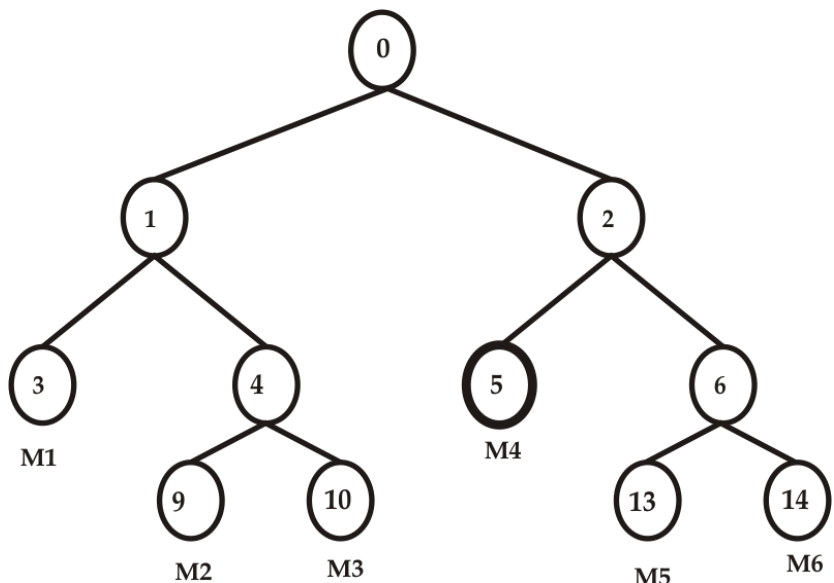

(a) Before member M8 join/Leave

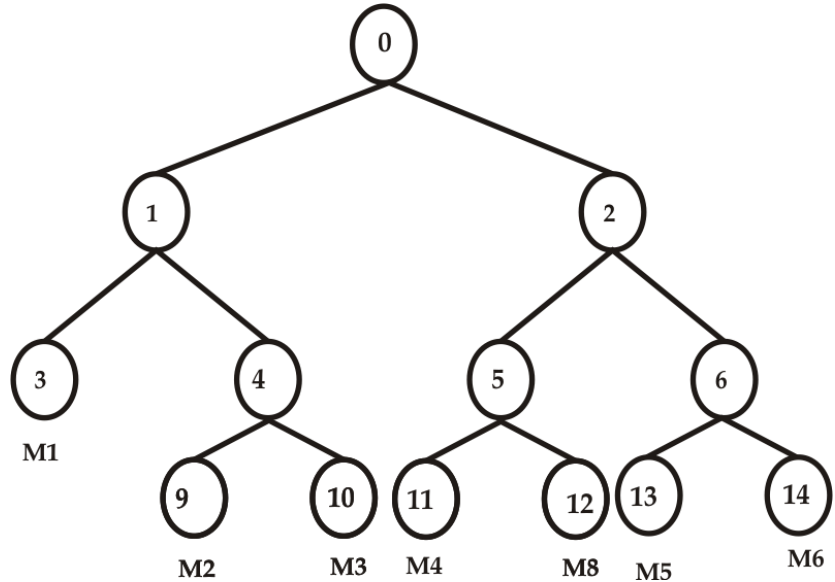

**(b) After M8 join/Leave the group**

**Fig.3. Member join/Leave case**

After joining, the support node becomes the responsible node to update the keys that are located in its path. It re-keys K5, K2, and K0 then broadcasts the public keys PK5 and PK2. The members M1, M2 and M3 compute K0 from the given PK2. Members M5 and M6 compute K2, K0 from the given public key PK5.

### 4.2  Member leaves

Assume that there are n members in the group currently where member $M_n$ leaves the group. Now, the support node becomes a responsible node to update the group key and to broadcast all the required public keys in the key tree. When a member leaves from the tree, its immediate left or right node will be uplifted higher by one level to reduce the number of keys to be updated by the support node. During the member leave operation, all the keys from the leaf node to the root node must be updated in order to prevent the access of future data by the left members from the group. This provides forward secrecy. If member $M_8$ wants to depart from the service, the internal node keys $K_5$, $K_2$ and $K_0$ must be renewed as shown in Fig.3b & Fig.3a. During the update phase, the support node $M_4$ becomes a responsible node to re-key the secret keys $K_2$ and $K_0$ and broadcasts the Public keys $PK_2$ and $PK_5$. The members $M_1$, $M_2$ and $M_3$ compute $K_0$ from the given $PK_2$. Members $M_5$ and $M_6$ compute $K_2$, $K_0$ from the given public key $PK_5$.

## 5  PERFORMANCE ANALYSIS

The proposed method has been implemented in JAVA for more than 500 users and we have analyzed the computation time with existing approaches to perform the rekeying operation. The BigInteger Java class was used for handling large numbers as key value in the key distribution protocol. The graphical result shown in Fig.4 is used to compare the group members key computation time that exists during the computation of group key of our proposed method with the existing methods. We evaluated key computation time for various distributed and collaborative key distribution protocols whose group sizes are considered starting from 128 to 576. The Tripartite Key Agreement protocol used in [14] is labeled as (TKA). The Diffie-Hellman based group keying algorithm used in this paper [16] is denoted as TGK (Tree-based Group Key agreement). The signature based secure group communication protocol proposed in the paper [17] is denoted as BMS (Bilinear pairing and Multi-Signature) in this paper. We implemented our proposed ETF (Euler Totient Function based) distributed key management algorithm to measure the computation time for various group sizes and compared the performance result with the various collaborative key distribution protocols and the result is shown in Fig.4. The implemented experimental graph shown below clearly depicts that the proposed Euler function based group keying approach has significant reduction in computation complexity. It is observed that when the group size is 512, the key computation time taken in ETF based approach by each group member is found to be 37ms, which is better in comparison with existing schemes.

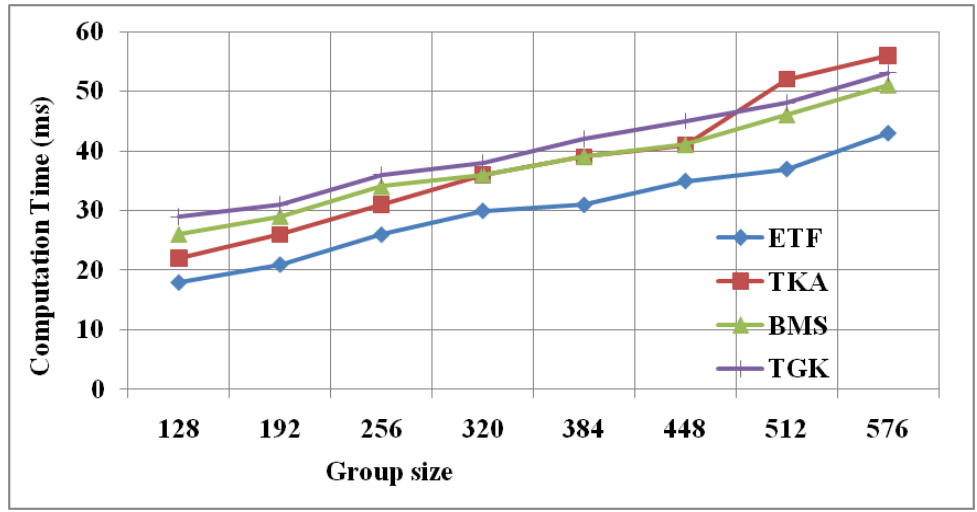

**Fig.4.Group member Key Computation Time for various Key Distribution methods**

## 6   ERROR CORRECTION USING LDPC CODES

This section discusses the error detection and correction methods used for detecting and correcting the errors that happens during the dissemination of keying information in distributed key management protocol.

### 6.1   Encoding at Sender

Encoding process at sender consists of three phases.

Phase 1: Conversion of original key information bits into binary values (0's and 1's)

Phase 2: Construction of Parity Check Matrix according to size of the key

Phase 3: Construction of Encoding stopping set

Phase 4: Generation and distribution of code words to group members

**Algorithm:**

Consider an example, where the size of key information is 8 bits. If the original key information bit is 8 bits [1 1 0 1 0 0 1 1] and its corresponding (8, 16) parity check matrix will be generated as mentioned in phase 2 and used as shown below. The group members are required to use the same size parity check matrix. From the parity check matrix the sender can construct the Tanner graph. The algorithm coverts the tanner graph into Pseudo tree based Encoding stopping set with maximum bit node degree 3 as explained in [18].

$$\begin{pmatrix} 1 & 0 & 0 & 0 & 0 & 0 & 1 & 0 & 0 & 0 & 0 & 0 & 1 & 0 & 0 \\ 0 & 1 & 1 & 1 & 1 & 0 & 0 & 0 & 1 & 0 & 0 & 1 & 1 & 1 & 0 & 0 \\ 1 & 0 & 1 & 0 & 0 & 1 & 1 & 0 & 0 & 1 & 1 & 1 & 0 & 0 & 0 & 0 \\ 1 & 0 & 0 & 1 & 1 & 0 & 0 & 1 & 0 & 1 & 1 & 0 & 1 & 0 & 1 & 1 \\ 0 & 0 & 0 & 0 & 1 & 0 & 1 & 1 & 1 & 0 & 0 & 0 & 0 & 1 & 0 \\ 0 & 0 & 0 & 0 & 0 & 1 & 0 & 1 & 0 & 1 & 0 & 0 & 0 & 1 & 0 & 1 \\ 0 & 1 & 0 & 0 & 0 & 0 & 0 & 1 & 0 & 1 & 0 & 1 & 0 & 0 & 1 \\ 0 & 1 & 1 & 1 & 0 & 1 & 0 & 0 & 0 & 0 & 1 & 0 & 0 & 1 & 0 \end{pmatrix}$$

*Parity Check Matrix of (8, 16) LDPC codes*

**Reevaluated bits:**
The reevaluated bits r1 and r2 are found in a twofold constraint encoding stopping set with key check nodes C7 and C8. The Key parity check equations for the check nodes C7 and C8 are computed by using Fig.6.

$$C7 = X11 \oplus X16$$
$$C8 = X4 \oplus X6 \oplus X12 \oplus X15$$

**Encoding Process:**
The stages of encoding are given below:
1. Fill the values of the information bits in the bottom most level,
 i.e., [X5 X6 X7 X10 X11 X12 X14 X15 ] = [1 1 0 1 0 0 1 1]. Initially assign X4 = 0 and X16 = 0.

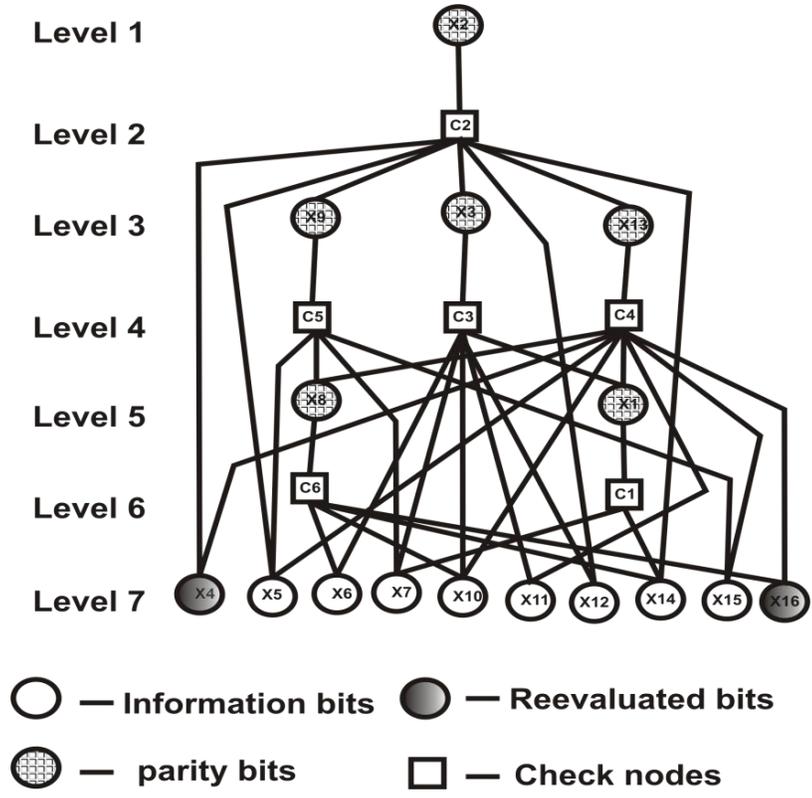

**Fig.5.** The Pseudo tree at sender

2. Encode the pseudo tree as shown in Fig.5 and compute the parity bits as follows
   $X8 = X6 \oplus X10 \oplus X14 \oplus X16 = 1$
   $X1 = X7 \oplus X14 = 1$
   $X9 = X5 \oplus X8 \oplus X7 \oplus X15 = 1$
   $X3 = X6 \oplus X1 \oplus X7 \oplus X10 \oplus X11 \oplus X12 = 0$
   $X13 = X4 \oplus X8 \oplus X1 \oplus X5 \oplus X10 \oplus X11 \oplus X15 \oplus X16 = 0$
   $X2 = X4 \oplus X9 \oplus X5 \oplus X3 \oplus X13 \oplus X12 \oplus X14 = 1$
3. Compute the values of key parity check equations C7 and C8 for the diagram shown in Fig.6.
   $C7 = X11 \oplus X9 \oplus X13 \oplus X16 \oplus X2 = 0$
   $C8 = X4 \oplus X2 \oplus X3 \oplus X6 \oplus X12 \oplus X15 = 1$
4. Since C7 = 0, C8 = 1, correct values of reevaluated bits X16 = 1, X4 = 0.
5. Compute all the parity bits again based on the new values of X16 and X4. The encoded code word is [X5 X6 X7 X10 X11 X12 X14 X15 X4 X16 X8 X1 X9 X3 X13 X2 ] = [1 1 0 1 0 0 1 1 0 1 0 1 0 1 1 0]

## 6.2 Decoding at Group Members side

Error correction process at each group member's area consists of three phases.
Phase 1: Receive the code words from the sender/support node.
Phase 2: Construction of encoding of stopping set as shown in Fig.6 according to the parity check matrix used by sender/support node.
Phase 3: Detection of errors by verifying the check node values.
Phase 4: Correction of errors.

**Decoding Process:**
The code word that is received from the sender is [X5 X6 X7 X10 X11 X12 X14 X15 X4 X16 X8 X1 X9 X3 X13 X2 ] = [1 1 0 1 0 0 1 1 0 1 0 1 0 1 1 0]. After the receipt of the code word, the group members should place the values in the encoding stopping set, and also should verify for the occurrence of errors. On encountering an error, the group members should find out the type of error where in the error can be a single bit error, two bit error,..., n bit error. For any type of errors, there are 3 cases available for the correction of errors and those cases are explained below.

**Case 1: Reevaluation bit**
The errors in this case are found to be in the reevaluated bit. For example, [X5 X6 X7 X10 X11 X12 X14 X15 X4 X16 X8 X1 X9 X3 X13 X2] = [1 1 0 1 0 0 1 1 0 <u>0</u> 0 1 0 1 1 0]. The reevaluated bit (X16) value 1 is changed to 0. This is a single bit error type. During the decoding process parity bit values are computed, and during such computation X16 = 0. This does not coincide with the received codeword, where the X16 bit value is 1. Hence, in such a scenario the single bit error has been found. After the conclusion of the occurrence of error, during the correction of errors the reevaluated bits are inverted as [X4  X16] = [0  1]. Even now if the error is not corrected then perform two bit error correction process. Again, the parity bit values are computed from bottom to top (i.e.) calculate all the parity bit values up to reach the last check nodes. If the key check node [C7 C8] values after parity computation are [0 0], then error has been rectified.

**Case 2: Key Information bits**
This is the case, where the error has been occurred in the original key information bits or aggregation of information bits and reevaluated bits. Considering the example of received code words as follows, [X5 X6 X7 X10 X11 X12 X14 X15 X4 X16 X8 X1 X9 X3 X13 X2 ] = [1 1 <u>1</u> 1 0 0 1 1 0 1 0 1 0 1 1 0], X7 information bit value 0 is changed to 1. During the decoding process some parity bit values will be changed and hence the key check node values will not end up as [0  0]. Hence, on finding such error, the correction follows the steps given below:

*Step 1:*
For correcting the error, all combination of reevaluated bits are changed and even after changing if the key check node [C7 C8] values does not become [0 0], migrate to step2.
*Step 2:*
This is the final stage of correction wherein, each information bit from left to right in the leaf node are changed until the key check nodes [C7 C8] value becomes [0 0], and the error has been rectified. There are also some cases, where even during such a change the key check node values may not become [0 0] and in such a situation, combination of two, three, …, n information bits are changed in leaf node from left to right to obtain the key check node value to be [0 0].

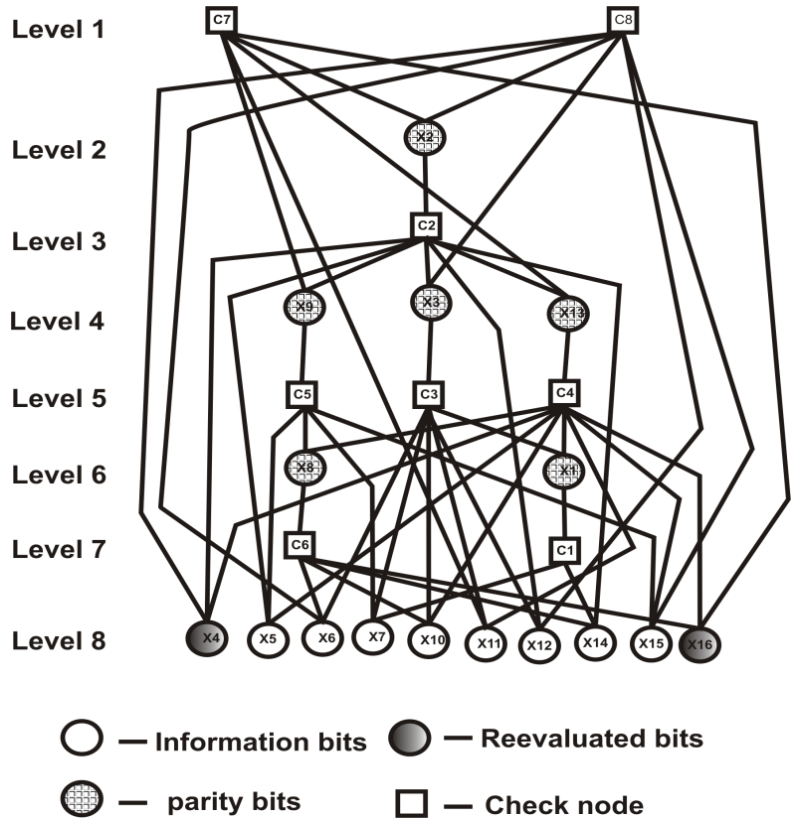

Fig.6. The Encoding Stopping set at SENDER

**Case 3: Parity bits**
In the second case, the error would have occurred in the parity bits: For example, considering the bits received are [X5 X6 X7 X10 X11 X12 X14 X15 X4 X16 X8 X1 X9 X3 X13 X2 ] = [1 1 0 1 0 0 1 1 0 1 0 1 <u>1</u> 1 1 0]. In this example X9 parity bit value 0 is changed to 1. During the decoding process while calculating the parity bit values, X9 = 0 will be obtained which is not in coincidence with the received codeword, where X9 bit value is 1. This error can be rectified automatically while correcting the information bits. The following proof gives the information regarding the number of changes for the different types of errors.
*Lemma:*
Any arbitrary LDPC codes has $O(n^2)$ time complexity during decoding process for n bit errors.
*Proof:*
    Let 's' be the number of leaf nodes which includes 'n' information bits and 'r' re-evaluated bits received from the sender. The received bits are substituted in the encoding stopping set generated at group member's side. Now we apply the decoding process in encoding stopping set.

An error is said to occur:
1. If the values of the level 1 check nodes (i.e., key check nodes) are not zero.
2. If the computed parity bit values and received parity bit values at each level, in encoding stopping set are unequal.

We need to correct these errors. In case, if the re-evaluated bits are corrupted, then complexity of correcting the re-evaluated bit is $O(3)$. Depending upon the number of encoding stopping sets the complexity may increase. If there are two encoding stopping set then the decoding time complexity is $O(6)$, in which $O(3)$ for first encoding stopping set and another $O(3)$ computation for second encoding stopping set and so on. On occurrence of error in the information bits, the following procedure has to be followed. Since the number of corrupted bits and their position are unknown, we correct them step by step procedure. First we change the 1st bit of the leaf nodes from left to right. Next, we compute the new parity bit values. If the key check node values are equal to zero, then the error is corrected.

Even now, if the key check node values are unequal to zero, then the second bit of the information bit is changed and the procedure is repeated until reaching the last information bits in the leaf level. From this it is very clear that the complexity for correcting one bit error is O(n). If still error persists, the above procedure is repeated for all combination of two information bits. Now the time complexity becomes O(n+(n(n-1)/2)). Even then if the error is uncorrected, then the combination of 'i' (i=3,4,…..,n) information bits are changed to calculate the new parity bit value, and the error is corrected. Hence the time complexity for the decoding procedure is $O(n^2)$ as follows. For example if the total number of received information bits is 4 bits and all the four information bits are corrupted, then the decoding time complexity can be computed as shown below.

$= n + (n ( n – 1) /2) +(( n – 1 ) ( n – 2 ) / 2 )+1$

$= n+((n^2 − n)/2)+( (n^2 − 3n + 2)/2)+1$

$= n^2 − n+2$

$= O(n^2)$

## 7    Concluding Remarks

In this paper, a new fully contributory binary tree based group key computation protocol for n bit numbers as the key value has been proposed for creating and distributing keys in order to provide effective security in group communications. The main advantages of our proposed approach are, the computation time takes place between GC and group members get reduced by using euler's totient function and it recovers the original keying information bits if the keying information's are corrupted. In order to do that we introduced two algorithms in this proposed work. First, combination of Euler's Totient function and diffie-hellman algorithm based key computation process. Second, Encoding stopping set is constructed in the sender and receiver side in order to verify whether the received key material has no errors. If any error is found in the received key at receiver side, decoding algorithm can correct the error in $O(n^2)$ time. The main advantage of this approach is that the proposed approach can correct n-bit errors in less decoding time. However the main concern of our proposed approach is that it also increases the storage complexity since the number of keys stored by group members is increased. Further extensions to this work are to devise techniques to reduce the storage complexity which is the amount of storage required to store the key related information in group member's area.